# Outer Bounds for the Interference Channel with a Cognitive Relay

Stefano Rini, Daniela Tuninetti, and Natasha Devroye
Department of Electrical and Computer Engineering
University of Illinois at Chicago
Email: {srini2, danielat, devroye}@uic.edu

*Abstract*—In this paper, we first present an outer bound for a general interference channel with a cognitive relay, i.e., a relay that has non-causal knowledge of both independent messages transmitted in the interference channel. This outer bound reduces to the capacity region of the deterministic broadcast channel and of the deterministic cognitive interference channel through nulling of certain channel inputs. It does not, however, reduce to that of certain deterministic interference channels for which capacity is known. As such, we subsequently tighten the bound for channels whose outputs satisfy an "invertibility" condition. This second outer bound now reduces to the capacity of this special class of deterministic interference channels. The second outer bound is further tightened for the high SNR deterministic approximation of the Gaussian interference channel with a cognitive relay by exploiting the special structure of the interference. We provide an example that suggests that this third bound is tight in at least some parameter regimes for the high SNR deterministic approximation of the Gaussian channel. Another example shows that the third bound is capacity in the special case where there are no direct links between the non-cognitive transmitters.

## I. Introduction

The interference channel with a cognitive relay (IFC-CR) is a channel model of contemporary interest as it encompasses several multi-user and cognitive channel models. The IFC-CR consists of a classical two-user interference channel in which the two independent messages are non-causally known at a third transmitter–termed the cognitive relay–who serves only to aid the other two in their transmissions. This five-node channel generalizes a number of known channels including the broadcast (BC), the interference (IFC), and the cognitive interference channel (C-IFC) as special cases.

**Past work.** The IFC-CR was first introduced in [1] and [2], where the message knowledge at the relay was obtained causally and non-causally, respectively. This channel model is also referred to as the "broadcast channel with cognitive relays" in [3]. In [1], [2] achievable rate regions that combine dirty-paper coding, beamforming and interference reduction techniques are derived for the Gaussian IFC-CR. In [4] the achievable rate region of [2] is improved upon and a sum-rate outer bound based on the MIMO Gaussian cognitive interference channel is proposed in order to determine the degrees of freedom of the Gaussian IFC-CR; it is shown that it is possible to achieve the degrees of freedom of a two user no-interference channel for a large range of channel parameters. In [3], the authors derive an achievable rate region that contains all previously known achievable rate regions. To the best of the authors' knowledge, outer bounds for a general (i.e., not Gaussian as in [4]) IFC-CR have not yet been considered.

**Contributions.** In this paper, we:
1) derive an outer bound for a general IFC-CR;
2) note that the derived outer bound reduces to the capacity region of deterministic BCs [5], of deterministic C-IFCs [6], but not to that of the class of deterministic IFCs studied in [7];
3) tighten it for deterministic IFC-CRs whose outputs satisfy an "invertibility" condition as in [7];
4) tighten it even further for the high-SNR linear deterministic approximation of the Gaussian IFC-CR (just referred to as *high-SNR channel* for short in the following), by generalizing the approach of [7], by exploiting the special interference structure;
5) illustrate the achievability of this last outer bound for some parameters of the high-SNR channel, possibly suggesting that the derived outer bound is tight in more parameter regimes.

**Organization.** The rest of the paper is organized as follows: Section II formally defines the channel model; Section III presents our outer bound, and shows how it may be tightened for certain deterministic IFC-CRs and for the high SNR channel; Section IV shows achievability of the tightened outer bound for certain parameters of the high SNR channel; and Section V concludes the paper.

## II. Channel model, notation and definitions

We consider the two-user IFC-CR depicted in Fig. 1, in which the transmission of the two independent messages $W_i \in \{1, 2, ..., 2^{NR_i}\}$, $i \in \{1, 2\}$, is aided by a single cognitive relay (whose input to the channel has subscript $c$). The relay is non-causally cognizant of both messages. We assume classical definitions for achievable rates, and capacity inner and outer bound regions [8]. The notation $P^N_{Y_1, Y_2|X_1, X_2, X_c}$ represents the $N$-fold memoryless extension of the channel $P_{Y_1, Y_2|X_1, X_2, X_c}$, which describes the relationship between the

The work of S. Rini and D. Tuninetti was partially funded by NSF under award 0643954.

channel inputs $X_1, X_2, X_c$ and the channel outputs $Y_1, Y_2$. The IFC-CR contains three well-studied multi-user channels as special cases:
a) Interference channel (IFC): if $X_c = \emptyset$;
b) Broadcast channel (BC): if $X_1 = X_2 = \emptyset$; and
c) Cognitive channel (C-IFC): if $X_1 = \emptyset$ or $X_2 = \emptyset$.

The largest known achievable rate region for the IFC-CR presented in [3] combines ideas from the achievable rate regions of these three special channel models it subsumes. In the next section we derive an outer bound for a general IFC-CR.

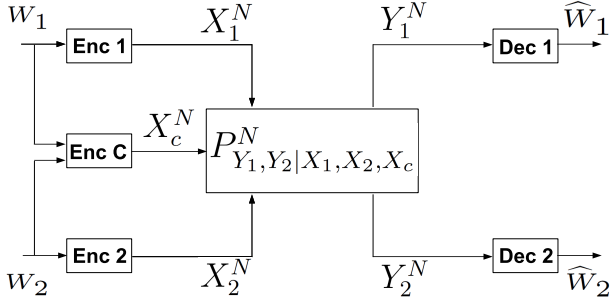

Fig. 1. The interference channel with a cognitive relay (IFC-CR).

## III. OUTER BOUNDS FOR THE IFC-CR

We first derive an outer bound valid for all memoryless IFC-CRs. We then tighten this bound by developing further inequalities for a class of deterministic channels and for the high SNR channel in the spirit of [7]. Finally, we evaluate our tightened bound for the high-SNR channel.

### A. General IFC-CR outer bounds

**Theorem III.1.** *If $(R_1, R_2)$ lies in the capacity region of the IFC-CR, then the following must hold for any $\widetilde{Y}_1$ and $\widetilde{Y}_2$ having the same marginal distributions as $Y_1$ and $Y_2$, respectively, but otherwise arbitrary correlated:*

$$R_1 \leq I(Y_1; X_1, X_c | Q, X_2), \tag{1a}$$
$$R_2 \leq I(Y_2; X_2, X_c | Q, X_1), \tag{1b}$$
$$R_1 + R_2 \leq I(Y_2; X_1, X_2, X_c | Q) + I(Y_1; X_1, X_c | Q, \widetilde{Y}_2, X_2), \tag{1c}$$
$$R_1 + R_2 \leq I(Y_1; X_1, X_2, X_c | Q) + I(Y_2; X_2, X_c | Q, \widetilde{Y}_1, X_1), \tag{1d}$$

*for some*
$$P_{Q,X_1,X_2,X_c,Y_1,Y_2} = P_Q P_{X_1|Q} P_{X_2|Q} P_{X_c|X_1,X_2,Q} P_{Y_1,Y_2|X_1,X_2,X_c}.$$

*Proof:* We only outline the proof here for sake of space. The proof of this theorem can be found in Appendix A.

The outer bound may be thought of as the intersection of two C-IFC outer bounds [6] obtained by non-causally providing (*genie*) one of the transmitters with the message of the other transmitter (as done in the [4] for the sum-rate of Gaussian IFC-CRs). For the sum-rates, since the receivers cannot cooperate, the capacity cannot depend on the correlation among the output signals, as first observed in [9] for BCs. By giving (genie side-information) a receiver a signal that has the same marginal distribution as the other user's output but that is arbitrarily correlated with its own output, we obtain the two sum-rate bounds. The same idea was used in [6] for the C-IFC and in [10] for cooperative IFCs. ∎

**Remark 1:** *The outer bound reduces to the capacity region of a deterministic BC when $X_1 = X_2 = \emptyset$ and to the capacity of a deterministic C-IFC when either $X_2 = \emptyset$ or $X_1 = \emptyset$. However, Th. III.1 does not reduce to the capacity region of the class of deterministic IFCs studied in [7] when $X_c = \emptyset$. In the following we thus develop additional rate bounds to cover this latter case.*

### B. Further bounds for a class of IFC-CRs

Consider, in the spirit of [7], IFC-CRs whose outputs satisfy:

$$Y_1 = f_1(X_1, X_c, V_{12}), \quad V_{12} = g_2(X_2, Z_1):$$
$$H(Y_1 | X_1, X_c) = H(V_{12} | X_1, X_c) = H(V_{12} | X_c), \tag{2a}$$
$$Y_2 = f_1(X_2, X_c, V_{21}), \quad V_{21} = g_1(X_1, Z_2):$$
$$H(Y_2 | X_2, X_c) = H(V_{21} | X_2, X_c) = H(V_{21} | X_c), \tag{2b}$$

where the functions $f_1, f_2, g_1$ and $g_2$ are deterministic, and $Z_1$ and $Z_2$ are "noise" random variables (RVs) independent of the inputs. Notice the invertibility conditions in (2a) and (2b) (and recall that $X_1 = X_1(W_1)$ is independent of $X_2 = X_2(W_2)$). We tighten the outer bound of Th.III.1 as follows:

**Theorem III.2.** *If $(R_1, R_2)$ lies in the capacity region of the IFC-CR, then the following must hold:*

$$R_1 \leq (1a), R_2 \leq (1b), R_1 + R_2 \leq \min\{(1c), (1d)\}, \tag{3a}$$
$$N(R_1 + R_2) \leq I(V_{21}^N; X_c^N) + H(Y_1^N | \widetilde{V}_{21}^N) - H(\widetilde{V}_{21}^N | X_1^N)$$
$$+ I(V_{12}^N; X_c^N) + H(Y_2^N | \widetilde{V}_{12}^N) - H(\widetilde{V}_{12}^N | X_2^N), \tag{3b}$$
$$N(2R_1 + R_2) \leq -H(\widetilde{V}_{21}^N | X_1^N) - 2H(V_{12}^N | X_2^N)$$
$$+ H(Y_1^N) + H(Y_1^N | \widetilde{V}_{21}^N, X_2^N) + H(Y_2^N | \widetilde{V}_{12}^N)$$
$$+ I(V_{12}^N; X_c^N) + I(V_{21}^N; X_c^N), \tag{3c}$$
$$N(R_1 + 2R_2) \leq -H(\widetilde{V}_{12}^N | X_2^N) - 2H(V_{21}^N | X_1^N)$$
$$+ H(Y_2^N) + H(Y_2^N | \widetilde{V}_{12}^N, X_1^N) + H(Y_1^N | \widetilde{V}_{21}^N)$$
$$+ I(V_{21}^N; X_c^N) + I(V_{12}^N; X_c^N), \tag{3d}$$

*where (3a) holds under the hypothesis of Th.III.1, and where $\widetilde{V}_{21}, \widetilde{V}_{12}$ are conditionally independent copies of $V_{12}$ and $V_{21}$, that is, distributed jointly with $(Q, X_1, X_2, X_c)$ with*

$$P_{\widetilde{V}_{21},\widetilde{V}_{12}|Q,X_1,X_2,X_c} = P_{\widetilde{V}_{21}|Q,X_1} P_{\widetilde{V}_{12}|Q,X_2}.$$

*Proof:* The proof may be found in Appendix B. ∎

**Remark 2:** *When $X_c = \emptyset$ the outer bound in Th. III.2 reduces to that of the class of deterministic IFCs considered in [7], which is tight for the high SNR IFC [7] and is to within one bit of a simple Han and Kobayashi achievable scheme for the Gaussian IFC [11].*

### C. Outer bound for the high SNR IFC-CR

The outer bound of Th. III.2 may be further tightened for the high-SNR IFC-CR. This channel, as developed in [7], models a Gaussian noise channel as the receive SNRs grow to infinity.

The high SNR channel is a deterministic binary linear channel with outputs:

$$Y_u = \boldsymbol{S}^{m-n_{u1}} X_1 \oplus \boldsymbol{S}^{m-n_{uc}} X_c \oplus \boldsymbol{S}^{m-n_{u2}} X_2, \quad (4)$$

for $u \in \{1, 2\}$, where the inputs are binary vectors of length $m \triangleq \max\{n_{11}, n_{12}, n_{21}, n_{22}, n_{1c}, n_{2c}\}$, $\boldsymbol{S}$ is a shift matrix of dimensions $m \times m$, and $\oplus$ denotes the binary XOR operation. The high SNR channel belongs to the class of deterministic IFC-CRs whose outputs are described by:

$$Y_1 = f_1(X_1, V_{1c}, V_{12}), \quad V_{12} = g_2(X_2), \quad V_{1c} = h_1(X_c),$$
$$Y_2 = f_2(X_2, V_{2c}, V_{21}), \quad V_{21} = g_1(X_1), \quad V_{2c} = h_2(X_c),$$

for some deterministic functions $f_1, f_2, g_1, g_2, h_1$ and $h_2$ and subject to the invertibility conditions in (2a) and (2b).

The capacity achieving strategy for the high SNR channel has provided insights on capacity approaching strategies for the corresponding Gaussian channel, and have allowed the derivation of capacity results to "within a constant gap" for the IFC [11] and C-IFC [12]. We hope that a similar result may be derived for the Gaussian IFC-CR using achievable schemes inspired by the high SNR approximation. For the high SNR channel, we tighten the rate bounds in (3) by replacing the term $I(V_{21}^N; X_c^N)$ with $I(V_{21}^N; V_{2c}^N)$, and the term $I(V_{12}^N; X_c^N)$ with $I(V_{12}^N; V_{1c}^N)$. This "substitution" of $X_c$ by $V_{2c}$ or $V_{1c}$ respectively is not possible in general since it is not generally known how the input $X_c$ affects the channel outputs. However, in the deterministic high SNR channel, the effect of the interference is specified by the deterministic functions $V_{1c} = h_1(X_c)$ and $V_{2c} = h_2(X_c)$.

**Remark 3:** *This step of tightening the bound highlights the stumbling block in deriving outer bounds for general IFC and BCs: in general we do not know the* exact *form of the interfering signal(s) at a given receiver for any possible input distribution. Assuming that the channel is deterministic and in a certain way "invertible", allows one to exactly determine the interference. Notice also that in the tightened bound, "conditioning" on the interference generated by $X_j$ at the output $Y_i$, given by $V_{ij}$ (rather than on $X_j$ itself), implies that the interference has been removed without necessarily decoding the message corresponding to $X_j$.*

Evaluation of the tightened bound for the high-SNR channel yields:

**Theorem III.3.** *If $(R_1, R_2)$ lies in the capacity region of the high SNR IFC-CR, then*

$$R_1 \leq \max\{n_{11}, n_{1c}\} \quad (5a)$$
$$R_2 \leq \max\{n_{22}, n_{2c}\} \quad (5b)$$
$$R_1 + R_2 \leq 1_{\{n_{11}-n_{1c} \neq n_{21}-n_{2c}\}} \left([n_{11} - \max\{n_{12}, n_{1c}\}]^+ \right.$$
$$+ \max\{n_{22} + n_{1c}, n_{2c} + n_{12}\})$$
$$+ 1_{\{n_{11}-n_{1c} = n_{21}-n_{2c}\}} (\max\{n_{22}, n_{21}, n_{2c}\}$$
$$+ [n_{11} - n_{21}]^+) \quad (5c)$$
$$R_1 + R_2 \leq 1_{\{n_{22}-n_{2c} \neq n_{12}-n_{1c}\}} \left([n_{22} - \max\{n_{21}, n_{2c}\}]^+ \right.$$
$$+ \max\{n_{11} + n_{2c}, n_{1c} + n_{21}\})$$
$$+ 1_{\{n_{22}-n_{2c} = n_{12}-n_{1c}\}} (\max\{n_{11}, n_{12}, n_{1c}\}$$
$$+ [n_{22} - n_{12}]^+) \quad (5d)$$
$$R_1 + R_2 \leq \max\{n_{11} - n_{21}, n_{12}, n_{1c}\} + \min\{n_{1c}, n_{12}\}$$
$$+ \max\{n_{22} - n_{12}, n_{21}, n_{2c}\} + \min\{n_{2c}, n_{21}\} \quad (5e)$$
$$2R_1 + R_2 \leq \max\{n_{11}, n_{12}, n_{1c}\}$$
$$+ \max\{n_{11} - n_{21}, n_{12}, n_{1c}\} + \min\{n_{1c}, n_{12}\}$$
$$+ \max\{n_{22} - n_{12}, n_{21}, n_{2c}\} + \min\{n_{2c}, n_{21}\} \quad (5f)$$
$$R_1 + 2R_2 \leq \max\{n_{22}, n_{21}, n_{2c}\}$$
$$+ \max\{n_{11} - n_{21}, n_{12}, n_{1c}\} + \min\{n_{1c}, n_{12}\}$$
$$+ \max\{n_{22} - n_{12}, n_{21}, n_{2c}\} + \min\{n_{2c}, n_{21}\}. \quad (5g)$$

*Proof:* The derivation of the rate region in (5) can be found in Appendix C. ∎

## IV. ACHIEVING THE OUTER BOUND IN TH. III.3

While it remains to be shown that the outer bound of Th.III.3 is tight for the general high SNR channel, in this section we demonstrate by example that it is achievable for certain channel parameters. We consider two examples: Example I: the *strong signal, mixed cognition and weak interference regime at both decoders* given by $n_{11} > n_{1c} > n_{12}$ and $n_{22} > n_{2c} > n_{21}$; and Example II: the *no-interference regime for both decoders* given by $n_{12} = n_{21} = 0$.

### A. Example I

**Corollary IV.1.** *In the case of strong signal, mixed cognition and weak interference at both decoders, the capacity is*

$$R_1 \leq n_{11},$$
$$R_2 \leq n_{22}.$$

*Proof:* It can be shown that the outer bound of Th.III.3 reduces to the region in Corollary IV.1 when $n_{11} > n_{1c} > n_{12}$ and $n_{22} > n_{2c} > n_{21}$. The formal proof of the achievability of the point $(R_1, R_2) = (n_{11}, n_{22})$ can be found in Appendix D. We provide a sketch of the proof aided by the graphical representation of the achievable scheme in Fig. 2. Our aim is to highlight the innovative cooperation strategy implemented by the cognitive relay compared to the capacity achieving strategies of the high SNR IFC [11] and of the high SNR C-IFC [13].

*Extensions of the IFC and C-IFC.* In Fig. 2, the left section represents the three channel inputs $X_1, X_c, X_2$ and the right section represents the channel outputs $Y_1$ and $Y_2$. Each output is the modulo-2 sum of the three (down-shifted) inputs. The blue blocks in the upper-left section are the bits sent by user 1; the red blocks in the lower-left section are the bits sent by user 2. The down-shifted version of blue and red blocks appear on the right section. When the cognitive relay is absent, our channel model reduces to the high SNR IFC of [11]. In this channel cooperation is not possible and the transmission of one encoder produces interference at the non intended receiver. Receiver 1 observes $n_{11} - n_{12}$ (blue) of the bits from encoder 1 above the $n_{12}$ (red) bits from encoder 2. Decoder 1 has no knowledge of this interference produced by encoder 2 and thus is able to decode only the most significant $n_{11} - n_{12}$ bits. Similarly, receiver 2 only decodes the most significant $n_{22} - n_{21}$ (red) bits received above the interference. Without cognitive relay is possible to achieve only $(R_1, R_2) = (n_{11} - n_{12}, n_{22} - n_{21})$. When the cognitive relay is present, it can pre-cancel the interference experienced by one decoder, as in the high SNR C-IFC [13]. Let the input of the cognitive relay be non-zero only in the blue shaded block. By placing in the blue shaded block the same $n_{21}$ (blue) bits that interfere at decoder 2, the relay pre-cancels the interference at this user. The achievable rates in this case are $(R_1, R_2) = (n_{11} - n_{12}, n_{22})$. In a similar manner the cognitive relay can pre-cancel the interference generated by user 2 at receiver 1 by using the red-shaded block in Fig. 2. With this strategy we are able to cancel the interference at a single decoder only.

*A unique scheme for the IFC-CR.* To achieve the outer bound $(R_1, R_2) = (n_{11}, n_{22})$, we must be able to pre-cancel the interference at both decoders simultaneously. To do so, let the cognitive transmitter send the sum of the inputs that grant the pre-cancelation of the interference at a single decoder, i.e., the XOR of the blue-shaded and of the red-shaded blocks in Fig. 2. With this input at the cognitive relay, $Y_1$ is the XOR of signal from transmitter 1 and a shifted version of the interference at decoder 2 (purple block). Decoder 1 is able to decode this set of bits since $n_{11} > n_{1c}$ and remove it from $Y_1$. Transmitter 2 operates in a similar manner by decoding a shifted version of the interference at receiver 1 and adding it to $Y_2$ to obtain the message transmitted by encoder 2. This shows that the rate point $(R_1, R_2) = (n_{11}, n_{22})$ is achievable. ∎

The cognitive relay effectively trades an unknown interference term with a known one that each receiver is able to decode. This strategy generalizes to the case when the pre-coding by the cognitive relay against the interference at one decoder may be decoded by the other. We are currently investigating the applicability of this idea in a more general setting.

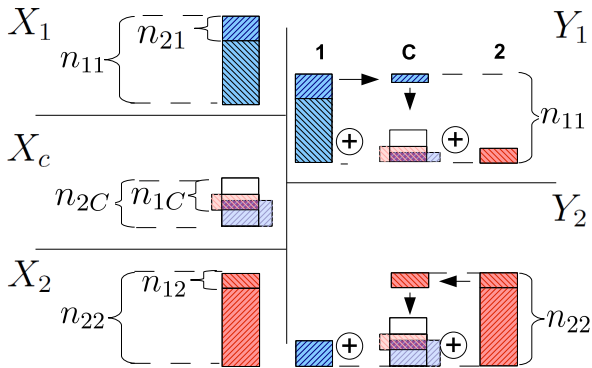

Fig. 2. Capacity achieving scheme for Example I.

### B. Example II

In this example we show that the outer bound of Th.III.3 is tight in the absence of interfering links: $n_{12} = n_{21} = 0$.

**Corollary IV.2.** *The capacity of the high SNR channel without interfering links is*

$$R_1 \leq \max\{n_{11}, n_{1c}\} \quad R_2 \leq \max\{n_{22}, n_{2c}\}$$
$$R_1 + R_2 \leq \max\{n_{22}, n_{2c}\} + \max\{n_{11}, n_{1c} - n_{2c}\}$$
$$R_1 + R_2 \leq \max\{n_{11}, n_{1c}\} + \max\{n_{22}, n_{2c} - n_{1c}\}.$$

**Remark 4:** *The region in Corollary IV.2 is a case where Th.III.1 and Th.III.2 coincide. In this case, if in addition either $n_{11} = 0$ or $n_{22} = 0$, the region reduces to the capacity region of the high SNR C-IFC determined in [6].*

*Proof:* It can be shown that the outer bound of Th.III.3 reduces to the region in Corollary IV.2 when $n_{11} > n_{1c} > n_{12}$ and $n_{22} > n_{2c} > n_{21}$.

We divide the achievability proof into three subcases. The achievability proofs of the first two cases below are deferred to Appendix E. The remaining achievability proof is presented graphically using the block representation introduced in Section IV-A. We note that all achievability proofs operate over a *single* channel use. All proofs are by inspection rather than through the systematic and judicious choice of RVs in a general achievable rate region such as that of [3]; a topic left for future work.

• **Capacity for weak cognition at both decoders:** when $n_{11} \geq n_{1c}$ and $n_{22} \geq n_{2c}$ the cognitive links $n_{1c}$ and $n_{2c}$ convey fewer clean bits than both direct links $n_{11}$ and $n_{22}$ respectively, and the outer bound reduces to $R_1 \leq n_{11}, R_2 \leq n_{22}$, achieved by keeping the cognitive relay silent.

• **Capacity for strong cognition at both decoders:** when $n_{11} < n_{1c}, n_{22} < n_{2c}$ the cognitive links $n_{1c}$ and $n_{2c}$ convey more clean bits than both direct links $n_{11}$ and $n_{22}$ respectively, and the outer bound simplifies to

$$R_1 \leq n_{1c}, \ R_2 \leq n_{2c}$$
$$R_1 + R_2 \leq \max\{n_{2c} + n_{11}, n_{1c}\}$$
$$R_1 + R_2 \leq \max\{n_{22} + n_{1c}, n_{2c}\}.$$

In Appendix E the corner points are achieved by having the two primary users send all the available clean bits along the respective direct links, and the cognitive relay utilizes its most significant bits to send bits for one user above the its direct link, attaining the single rate bound, and may use (parts of) its least significant bits to convey clean bits to the other user without creating interference with the direct transmissions.

• **Capacity for strong cognition for one decoder and weak cognition at the other:** when $n_{11} \geq n_{1c}, n_{22} < n_{2c}$ the cognitive link $n_{1c}$ conveys less clean bits to decoder 1 than the direct link $n_{11}$; the reverse is true for $n_{2c}$ and $n_{22}$. The condition $n_{11} < n_{1c}, n_{22} \geq n_{2c}$ is obtained by switching the role of the users. In this case, the outer bound becomes:

$$R_1 \leq n_{11}, \ R_2 \leq n_{2c}$$
$$R_1 + R_2 \leq n_{11} + \max\{n_{22}, n_{2c} - n_{1c}\}.$$

We again try to achieve the two corner points, but in this case each requires a different achievability scheme. We denote the binary vector of $R_{iP}$ bits for user $i$ as $\boldsymbol{b}_i^{R_{1P}}$. Similarly $(\boldsymbol{b}_i)_k^j$ indicates the bits between position $k$ and $j$ of $\boldsymbol{b}_i$. We use $\boldsymbol{0}^j$ to indicate a vector of length $j$ of all zeros.

*Corner point 1:* To achieve the corner point where the rate bound for $R_1$ meets the sum rate outer bound,
i) TX 1: transmits $n_{11}$ bits to RX 1 as

$$X_1 = [\boldsymbol{b}_1^{n_{11}}\ \mathbf{0}^{m-n_{11}}]^T,$$

where $A^T$ denotes the transpose of vector $A$,
ii) TX c: the cognitive relay transmits $[n_{2c}-n_{1c}]^+$ bits in the least significant bits from the cognitive relay to RX 2 without creating interference at RX 1 as

$$X_c = [\boldsymbol{b}_2^{[n_{2c}-n_{1c}]^+}\ \mathbf{0}^{m-[n_{2c}-n_{1c}]^+}]^T,$$

iii) TX 2: transmits $[n_{22}-[n_{2c}-n_{1c}]^+]^+$ to be received above the bits broadcasted from the cognitive relay at RX 2 as

$$X_2 = [(\boldsymbol{b}_2)_{[n_{2c}-n_{1c}]^+}^{[n_{2c}-n_{1c}]^++[n_{22}-[n_{2c}-n_{1c}]^+]^+}\ \mathbf{0}^{m-[n_{22}-[n_{2c}-n_{1c}]^+]^+}]^T.$$

Fig. 3 graphically illustrates the scheme.

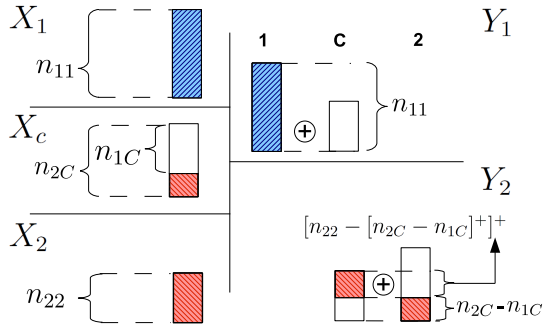

Fig. 3. The case of strong cognition for one decoder and weak cognition at the other, achievability scheme for corner point 1.

*Corner point 2:* To achieve the corner point where the rate bound for $R_2$ meets the sum rate outer bound, we again use a similar strategy but this time TX 1 sends additional bits above the interference coming from the cognitive relay, as:
i) TX 2 transmits $n_{22}$ bits for RX 2 through the direct link

$$X_2 = [\boldsymbol{b}_2^{n_{22}}\ \mathbf{0}^{m-n_{22}}]^T,$$

ii) the cognitive relay transmits $n_{2c}-n_{22}$ bits in the most significant bits for RX 2 achieving the full rate $R_2 = n_{2c}$ as

$$X_c = [\boldsymbol{b}_2^{n_{2c}-n_{22}}\ \mathbf{0}^{m-n_{2c}+n_{22}}]^T,$$

iii) TX 1 transmits $[n_{1c}-n_{2c}+n_{22}]^+$ below the interference from the cognitive relay at RX 1. It also transmits $n_{11}-n_{1c}$ bits that will be received above the interference from the cognitive relay as

$$\begin{aligned}X_1 &= [\boldsymbol{b}_1^{n_{11}-n_{1c}}\ \mathbf{0}^{\max\{n_{1c},n_{2c}-n_{22}\}}\\ &\quad (\boldsymbol{b}_1)_{n_{11}-n_{1c}}^{n_{11}-n_{1c}+[n_{1c}-n_{2c}+n_{22}]^+}\ \mathbf{0}^{m-n_{11}}]^T.\end{aligned}$$

Fig. 4 graphically illustrates the scheme. ∎

## V. CONCLUSIONS

In this work we derived the first general outer bounds for the interference channel with a cognitive relay and showed the achievability of the proposed outer bound for the high SNR deterministic approximation of the Gaussian interference channel with a cognitive relay for certain parameter regimes. The proposed outer bound is also tight for the deterministic channel models it encompasses: the deterministic broadcast channel, certain deterministic interference channels, and the deterministic cognitive interference channel. Our results leave multiple interesting open questions. We are currently investigating whether the presented high SNR outer bound is tight in all parameter regimes. Outcomes and insights obtained for the general high SNR capacity region will be then used to possibly determine a constant gap between an inner and our outer bound for the Gaussian channel–a result that would generalize numerous "constant gap" results including that of the interference and cognitive interference channels.

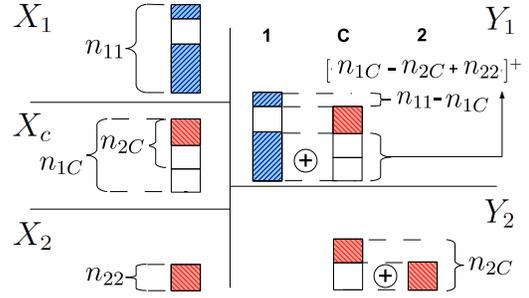

Fig. 4. The case of strong cognition for one decoder and weak cognition at the other, achievability scheme for corner point 2.

APPENDIX

*A. Proof of Theorem III.1*

By Fano's inequality, $H(W_i|Y_i^N) \leq N\epsilon_N$ for $P_e \to 0$, we can write

$$\begin{aligned}
N(R_1 - \epsilon_N) &\leq I(W_1; Y_1^N) \\
&\leq I(W_1; Y_1^N, W_2) \\
&= I(W_1; Y_1^N|W_2) \\
&= H(Y_1^N|W_2) - H(Y_1^N|W_1, W_2) \\
&= H(Y_1^N|W_2, X_2^N(W_2)) - H(Y_1^N|W_1, W_2, X_1^N(W_1), X_2^N(W_2), X_c^N(W_1, W_2)) \\
&\leq H(Y_1^N|X_2^N) - H(Y_1^N|X_1^N, X_2^N, X_c^N) \\
&= I(Y_1^N; X_1^N, X_c^N|X_2^N) \\
&\leq N(I(Y_1; X_1, X_c|X_2, Q)),
\end{aligned}$$

where $Q$ is a time sharing RV independent from all the other RVs and uniformly distributed on $\{1...N\}$. Similarly for user 2:

$$R_2 - \epsilon_N \leq I(Y_2; X_2, X_c|X_1, Q).$$

Next, let $\widetilde{Y}_u^N$ have the same marginal distribution as $Y_u^N$, $u \in \{1, 2\}$. We obtain the following Sato-type bounds [9]:

$$\begin{aligned}
N(R_1 + R_2 - 2\epsilon_N) &\leq I(Y_1^N; W_1) + I(Y_2^N; W_2) \\
&\leq I(Y_1^N; W_1|W_2) + I(Y_2^N; W_2) \\
&\leq I(Y_1^N, \widetilde{Y}_2^N; W_1|W_2) + I(Y_2^N; W_2) \\
&= H(Y_2^N) - H(\widetilde{Y}_2^N|W_1, W_2) + H(Y_1^N|\widetilde{Y}_2^N, W_2) - H(Y_1^N|\widetilde{Y}_2^N, W_2, W_1) \\
&\leq I(Y_2^N; X_1^N, X_2^N, X_c^N) + I(Y_1^N; X_1^N, X_c^N|\widetilde{Y}_2^N, X_2^N) \\
&\leq N(I(Y_2; X_1, X_2, X_c|Q) + I(Y_1; X_1, X_c|Q, \widetilde{Y}_2, X_2)).
\end{aligned}$$

By swapping the role of the users another sum-rate bound is obtained:

$$R_1 + R_2 - 2\epsilon_N \leq I(Y_1; X_1, X_2, X_c|Q) + I(Y_2; X_2, X_c|Q, \widetilde{Y}_1, X_1).$$

*Deterministic case:* For the deterministic IFC-CR of Section III-B, region (1) reduces to

$$\begin{aligned}
R_1 &\leq H(Y_1|X_2) & (6a) \\
R_2 &\leq H(Y_2|X_1) & (6b) \\
R_1 + R_2 &\leq H(Y_2) + H(Y_1|Y_2, X_2) & (6c) \\
R_1 + R_2 &\leq H(Y_1) + H(Y_2|Y_1, X_1). & (6d)
\end{aligned}$$

This is readily obtained by applying condition (2) to region (1) and letting $Y_i = \widetilde{Y}_i$, $i \in \{1, 2\}$.

*B. Proof of Theorem III.2*

Given the random variables $(Q, X_1, X_2, X_c)$ with probability distribution $P_{Q,X_1,X_2,X_c} = P_Q P_{X_1|Q} P_{X_2|Q} P_{X_c|Q,X_1,X_2}$, let $\widetilde{V}_{21}$ and $\widetilde{V}_{12}$ be conditionally independent copies of $V_{21}$ and $V_{12}$, that is, distributed jointly with $(Q, X_1, X_2, X_c)$ as $P_{\widetilde{V}_{21},\widetilde{V}_{12}|Q,X_1,X_2,X_c} = P_{\widetilde{V}_{21}|Q,X_1} P_{\widetilde{V}_{12}|Q,X_2}$. Similar arguments to those in [7] yield:

$$\begin{aligned}
N(R_1 + R_2 - 2\epsilon_N) &\leq I(W_1; Y_1^N, \widetilde{V}_{21}^N) + I(W_2; Y_2^N, \widetilde{V}_{12}^N) \\
&= H(\widetilde{V}_{21}^N) - H(\widetilde{V}_{21}^N|W_1, X_1^N) + H(Y_1^N|\widetilde{V}_{21}^N) - H(Y_1^N|\widetilde{V}_{21}^N, W_1, X_1^N) \\
&\quad + H(\widetilde{V}_{12}^N) - H(\widetilde{V}_{12}^N|W_2, X_2^N) + H(Y_2^N|\widetilde{V}_{12}^N) - H(Y_2^N|\widetilde{V}_{12}^N, W_2, X_2^N) \\
&\stackrel{(a)}{\leq} H(\widetilde{V}_{21}^N) - H(\widetilde{V}_{21}^N|X_1^N) + H(Y_1^N|\widetilde{V}_{21}^N) - H(Y_1^N|\widetilde{V}_{21}^N, W_1, X_1^N, X_c^N) \\
&\quad + H(\widetilde{V}_{12}^N) - H(\widetilde{V}_{12}^N|X_2^N) + H(Y_2^N|\widetilde{V}_{12}^N) - H(Y_2^N|\widetilde{V}_{12}^N, W_2, X_2^N, X_c^N) \\
&\stackrel{(b)}{=} H(Y_1^N|\widetilde{V}_{21}^N) + H(Y_2^N|\widetilde{V}_{12}^N) - H(\widetilde{V}_{21}^N|X_1^N) - H(\widetilde{V}_{12}^N|X_2^N) + H(\widetilde{V}_{21}^N) - H(V_{21}^N|\widetilde{V}_{12}^N, X_2^N, X_c^N) \\
&\quad + H(\widetilde{V}_{12}^N) - H(V_{12}^N|\widetilde{V}_{21}^N, X_1^N, X_c^N) \\
&\stackrel{(c)}{=} H(Y_1^N|\widetilde{V}_{21}^N) + H(Y_2^N|\widetilde{V}_{12}^N) - H(\widetilde{V}_{21}^N|X_1^N) - H(\widetilde{V}_{12}^N|X_2^N) + H(\widetilde{V}_{21}^N) - H(V_{21}^N|X_c^N) \\
&\quad + H(\widetilde{V}_{12}^N) - H(V_{12}^N|X_c^N) \\
&= I(V_{21}^N; X_c^N) + H(Y_1^N|\widetilde{V}_{21}^N) - H(\widetilde{V}_{21}^N|X_1^N) + I(V_{12}^N; X_c^N) + H(Y_2^N|\widetilde{V}_{12}^N) - H(\widetilde{V}_{12}^N|X_2^N),
\end{aligned}$$

where the inequality in "(a)" follows from further conditioning on $X_c$, the equality in "(b)" follows from the the assumed determinism, the equality in "(c)" follows from the conditional independence of the side information $\widetilde{V}_{ij}$ and the fact that $\widetilde{V}_{ij} = \widetilde{V}_{ij}(X_j, \widetilde{N}_i) \perp (X_i, N_j)$.

Similarly we have

$$\begin{aligned}
N(2R_1 + R_2 - 3\epsilon_N) &\leq I(W_1; Y_1^N, \widetilde{V}_{21}^N | W_2) + I(W_1; Y_1^N) + I(W_2; Y_2^N, \widetilde{V}_{12}^N) \\
&= H(Y_1^N | W_2, \widetilde{V}_{21}^N, X_2^N) - H(Y_1^N | W_1, W_2, \widetilde{V}_{21}^N, X_1^N, X_2^N, X_c^N) \\
&\quad + H(Y_1^N) - H(Y_1^N | W_1, X_1^N) \\
&\quad + H(Y_2^N | \widetilde{V}_{12}^N) - H(Y_2^N | W_2, \widetilde{V}_{12}^N, X_2^N) \\
&\quad + H(\widetilde{V}_{21}^N | W_2, X_2^N) - H(\widetilde{V}_{21}^N | W_1, W_2, X_1^N, X_2^N, X_c^N) \\
&\quad + H(\widetilde{V}_{12}^N) - H(\widetilde{V}_{12}^N | W_2, X_2^N) \\
&\stackrel{(a)}{\leq} H(Y_1^N | \widetilde{V}_{21}^N, X_2^N) - H(Y_1^N | \widetilde{V}_{21}^N, X_1^N, X_2^N, X_c^N) \\
&\quad + H(Y_1^N) - H(Y_1^N | W_1, X_1^N, X_c^N) \\
&\quad + H(Y_2^N | \widetilde{V}_{12}^N) - H(Y_2^N | W_2, \widetilde{V}_{12}^N, X_2^N, X_c^N) \\
&\quad + H(\widetilde{V}_{21}^N) - H(\widetilde{V}_{21}^N | X_1^N) \\
&\quad + H(\widetilde{V}_{12}^N) - H(\widetilde{V}_{12}^N | X_2^N) \\
&\stackrel{(b)}{=} H(\widetilde{V}_{21}^N) - H(\widetilde{V}_{21}^N | X_1^N) \\
&\quad + H(Y_1^N | \widetilde{V}_{21}^N, X_2^N) - H(V_{12}^N | \widetilde{V}_{21}^N, X_1^N, X_2^N, X_c^N) \\
&\quad + H(Y_1^N) - H(V_{12}^N | X_1^N, X_c^N) \\
&\quad + H(\widetilde{V}_{12}^N) - H(\widetilde{V}_{12}^N | X_2^N) \\
&\quad + H(Y_2^N | \widetilde{V}_{12}^N) - H(V_{21}^N | \widetilde{V}_{12}^N, X_2^N, X_c^N) \\
&\stackrel{(c)}{=} H(\widetilde{V}_{21}^N) - H(\widetilde{V}_{21}^N | X_1^N) \\
&\quad + H(Y_1^N | \widetilde{V}_{21}^N, X_2^N) - H(V_{12}^N | X_2^N) \\
&\quad + H(Y_1^N) - H(V_{12}^N | X_c^N) \\
&\quad + H(\widetilde{V}_{12}^N) - H(\widetilde{V}_{12}^N | X_2^N) \\
&\quad + H(Y_2^N | \widetilde{V}_{12}^N) - H(V_{21}^N | X_c^N) \\
&= H(Y_1^N) + H(Y_1^N | \widetilde{V}_{21}^N, X_2^N) + H(Y_2^N | \widetilde{V}_{12}^N) \\
&\quad + I(V_{12}^N; X_c^N) + I(V_{21}^N; X_c^N) - H(\widetilde{V}_{21}^N | X_1^N) - 2H(V_{12}^N | X_2^N),
\end{aligned}$$

where again the inequality in "(a)" follows from further conditioning on $X_c$, the equality in "(b)" follows from the assumed determinism, the equality in "(c)" follows from the conditional independence of the side information $\widetilde{V}_{ij}$ and the fact that $\widetilde{V}_{ij} = \widetilde{V}_{ij}(X_j, \widetilde{N}_i) \perp (X_i, N_j)$.

We note here that the single-letterization of this bound is not straightforward. For instance consider the term $I(V_{ij}^N; X_c^N)$ $i, j \in \{1, 2\}$ $i \neq j$, for which we write

$$\begin{aligned}
I(V_{ij}^N; X_c^N) &= \sum_{k=1}^{N} \left( H\left((V_{ij})_k | (V_{ij})^{k-1}\right) - H\left((V_{ij})_k | (V_{ij})^{k-1}, X_c^N\right) \right) \\
&\leq \sum_{k=1}^{N} \left( H\left((V_{ij})_k\right) - H\left((V_{ij})_k | (V_{ij})^{k-1}, X_c^N\right) \right)
\end{aligned}$$

Since now $V_{ij}^N = g_i(X_j^N)$, it is not possible to drop the term $(V_{ij})^{k-1}, (X_c)^{k-1}, (X_c)_{k+1}^N)$ from the conditioning of the second term. Yet it is still possible to upper bounded this term as

$$I(V_{ij}^N; X_c^N) \leq \min\{H(V_{ij}^N), H(X_c^N)\}$$

and then proceed with the single letterization of the individual entropy expressions. This argument is used in deriving the bound of Theorem III.3.

The remaining bounds are by swapping the role of the users.

*C. Proof of Theorem III.3*

We show how expression (1) reduces to (5) in the high SNR IFC-CR. Given the symmetry of the channel, this equality can be established by showing the following set of equalities:

$$\begin{aligned}
(1a) &= (5a) \\
(1c) &= (5d) \\
(3b) &= (5e) \\
(3c) &= (5f),
\end{aligned}$$

since switching the role of the users, we obtain the equalities

$$(1b) = (5b)$$
$$(1d) = (5c)$$
$$(3d) = (5g).$$

Since $Y_1$ and $Y_2$ are binary vectors, their entropy is maximized when each component $(Y_i)_j$, $i \in \{1,2\}$, $j \in \{1...\max\{n_{ik}\}\}$, $k \in \{1,c,2\}$ has Bernoulli distribution with $p = 1/2$ ($\mathcal{B}(1/2)$). The remaining channel outputs are always zero because of the down-shifting operation and are of no interest. This output distribution can be achieved when all the inputs $(X_k)_j$, $k \in \{1,c,2\}$, $j \in \{1...n_{ik}\}$, $i \in \{1,2\}$ are iid $\mathcal{B}(1/2)$ since the modulo 2 sum of independent Bernoulli RVs is an independent Bernoulli RV.

For the first equation we have

$$\begin{aligned}(1a) &= H(Y_1|X_2) \\ &= H(\boldsymbol{S}^{m-n_{1c}}X_c \oplus \boldsymbol{S}^{m-n_{11}}X_1) \\ &\leq \max\{n_{1c}, n_{11}\} = (5a),\end{aligned}$$

where the last passage follows from the fact that $\boldsymbol{S}^{m-n_{1c}}X_c \oplus \boldsymbol{S}^{m-n_{11}}X_1$ is a binary vector with non zero components in the index interval $[1, m - \max\{n_{1c}, n_{11}\}]$. Equality in the last passage is achieved when $X_c$ and $X_1$ are vectors of iid $\mathcal{B}(1/2)$ components. Similarly:

$$\begin{aligned}(1c) &= H(Y_2) + H(Y_1|X_2, Y_2) \\ &= H(\boldsymbol{S}^{m-n_{21}}X_1 \oplus \boldsymbol{S}^{m-n_{22}}X_2 \oplus \boldsymbol{S}^{m-n_{2c}}X_c) + H(\boldsymbol{S}^{m-n_{11}}X_1 \oplus \boldsymbol{S}^{m-n_{1c}}X_c|\boldsymbol{S}^{m-n_{21}}X_1 \oplus \boldsymbol{S}^{m-n_{2c}}X_c) \\ &\leq \max\{n_{22}, n_{21}, n_{2c}\} \\ &\quad + 1_{\{n_{11}-n_{1c}\neq n_{21}-n_{2c}\}} \max\{n_{11} - [n_{21} - n_{2c}]^+, n_{1c} - [n_{2c} - n_{21}]^+\} \\ &\quad + 1_{\{n_{11}-n_{1c}=n_{21}-n_{2c}\}} [n_{11} - n_{21}]^+\end{aligned}$$

Consider the case $n_{11} - n_{1c} \neq n_{21} - n_{2c}$ and $n_{21} \geq n_{2c}$: the expression is now simplified as

$$\begin{aligned}&\max\{n_{22}, n_{21}\} + \max\{n_{11} - n_{21} + n_{2c}, n_{1c}\} = \\ &\max\{n_{22} + n_{11} - n_{21} + n_{2c}, n_{22} + n_{1c}, n_{11} + n_{2c}, n_{21} + n_{1c}\} = \\ &\max\{n_{22} - n_{21} + (n_{11} + n_{2c}), n_{22} - n_{21} + (n_{21} + n_{1c}), n_{11} + n_{2c}, n_{21} + n_{1c}\} = \\ &[n_{22} - n_{21}]^+ + \max\{n_{11} + n_{2c}, n_{1c} + n_{21}\}.\end{aligned}$$

On the other hand, when $n_{11} - n_{1c} \neq n_{21} - n_{2c}$ and $n_{21} < n_{2c}$ we have

$$\begin{aligned}&\max\{n_{22}, n_{2c}\} + \max\{n_{11}, n_{1c} - n_{2c} + n_{21}\} = \\ &\max\{n_{22} + n_{11}, n_{22} + n_{1c} - n_{2c} + n_{21}, n_{2c} + n_{11}, n_{1c} + n_{21}\} = \\ &\max\{n_{22} - n_{2c} + (n_{2c} + n_{11}), n_{22} - n_{2c} + (n_{21} + n_{1c}), n_{2c} + n_{11}, n_{1c} + n_{21}\} = \\ &[n_{22} - n_{2c}]^+ + \max\{n_{11} + n_{2c}, n_{1c} + n_{21}\}.\end{aligned}$$

Therefore we rewrite the case $n_{11} - n_{1c} \neq n_{21} - n_{2c}$ in a compact form as

$$[n_{22} - \max\{n_{21}, n_{2c}\}]^+ + \max\{n_{11} + n_{2c}, n_{1c} + n_{21}\}$$

With this last simplification we obtain equation (5d).

When letting $\widetilde{V}_{21} = V_{21}$ in equation (3b) we obtain

$$\begin{aligned}(3b) &= H(Y_1|V_{21}) + H(Y_2|V_{12}) + H(V_{21}) - H(V_{21}|X_c) + H(V_{12}) - H(V_{12}|X_c) \\ &= \max\{n_{11} - n_{21}, n_{12}, n_{1c}\} + \max\{n_{21}, n_{22} - n_{12}, n_{2c}\} + n_{12} - [n_{12} - n_{1c}]^+ + n_{21} - [n_{21} - n_{2c}]^+ \\ &= \max\{n_{11} - n_{21}, n_{12}, n_{1c}\} + \min\{n_{1c}, n_{12}\} + \max\{n_{22} - n_{12}, n_{21}, n_{2c}\} + \min\{n_{2c}, n_{21}\} = (5e).\end{aligned}$$

Finally, in the last case, again let $\widetilde{V}_{21} = V_{21}$ and $\widetilde{V}_{12} = V_{12}$

$$\begin{aligned}(3c) &= H(Y_1) + H(Y_2|V_{12}) + H(Y_1|V_{21}) + I(V_{12}^N; X_c^N) + I(V_{21}^N; X_c^N) \\ &= \max\{n_{11}, n_{12}, n_{1c}\} + \max\{n_{21}, n_{22} - n_{12}, n_{2c}\} + \max\{n_{11} - n_{21}, n_{12}, n_{1c}\} \\ &\quad + \min\{n_{1c}, n_{12}\} + \min\{n_{2c}, n_{21}\} = (5f).\end{aligned}$$

This concludes the proof.

## D. Proof of Corollary IV.1

We consider the case of strong signal, mixed cognition and weak interference at both decoders, that is

$$n_{11} > n_{1c} > n_{12} \tag{7a}$$
$$n_{22} > n_{2c} > n_{21}. \tag{7b}$$

In this case the outer bound reads

$$\begin{aligned}
R_1 &\leq n_{11} \\
R_2 &\leq n_{22} \\
R_1 + R_2 &\leq n_{11} - n_{12} + n_{22} + n_{12} \\
R_1 + R_2 &\leq n_{22} - n_{2c} + n_{11} - n_{2c} \\
R_1 + R_2 &\leq \max\{n_{11} - n_{21}, n_{1c}\} + \max\{n_{22} - n_{12}, n_{2c}\} + n_{12} + n_{21} \\
2R_1 + R_2 &\leq \max\{n_{11} - n_{21}, n_{1c}\} + \max\{n_{22} - n_{12}, n_{2c}\} + n_{12} + n_{21} + n_{11} \\
R_1 + 2R_2 &\leq \max\{n_{22} - n_{12}, n_{2c}\} + \max\{n_{11} - n_{21}, n_{1c}\} + n_{12} + n_{21} + n_{22}.
\end{aligned}$$

From the conditions (7), it further simplifies as:

$$\begin{aligned}
R_1 &\leq n_{11} \\
R_2 &\leq n_{22} \\
R_1 + R_2 &\leq n_{11} + n_{22} \\
R_1 + R_2 &\leq \max\{n_{11} - n_{21}, n_{1c}\} + \max\{n_{22} - n_{12}, n_{2c}\} + n_{12} + n_{21}.
\end{aligned}$$

To show that the last sum rate is loose we write

$$\max\{n_{11} - n_{21}, n_{1c}\} + \max\{n_{22} - n_{12}, n_{2c}\} + n_{12} + n_{21} = $$
$$\max\{n_{11} + n_{12}, n_{1c} + n_{12} + n_{21}\} + \max\{n_{22} + n_{21}, n_{2c}\} \geq n_{11} + n_{22}.$$

From this last observation we conclude that the outer bound of Th. III.3 reduces to

$$R_1 \leq n_{11}$$
$$R_2 \leq n_{22}.$$

We now prove the achievability of this outer bound. Since the outer bound expression describes a rectangle, we need to show the achievability of the corner point $(R_1, R_2) = (n_{11}, n_{22})$: time sharing assures the achievability of the whole region when this corner point is achievable. The achievable scheme is described in Section IV-A: both transmitters sends $n_{ii}$, $i \in [1,2]$ to the respective decoders while the cognitive relay pre-cancels the interference at both decoders. This pre-cancelation introduces an additional interference that the decoders are able to decode and eliminate.

Consider the following transmission scheme:

i) TX 1 transmits $n_{11}$ bits for RX 1 through the direct link:

$$X_1 = [\ \boldsymbol{b}_1^{n_{11}}\ \boldsymbol{0}^{m-n_{11}}]^T,$$

ii) TX 2 transmits $n_{22}$ bits for RX 2 through the direct link:

$$X_2 = [\boldsymbol{b}_2^{n_{22}}\ \boldsymbol{0}^{m-n_{22}}]^T,$$

iii) to cancel the interference experienced at receiver 1, the relay produces:

$$X_c^{(1)} = [\boldsymbol{0}^{n_{1c}-n_{12}}\ (\boldsymbol{b}_2)_{n_{22}-n_{12}}^{n_{22}}\ \boldsymbol{0}^{m-n_{1c}}]^T.$$

Similarly to cancel the interference at receiver 2, the relay produces:

$$X_c^{(2)} = [\boldsymbol{0}^{n_{2c}-n_{21}}\ (\boldsymbol{b}_1)_{n_{11}-n_{21}}^{n_{11}}\ \boldsymbol{0}^{m-n_{2c}}]^T.$$

As explained is Section IV-A, the strategy that achieves capacity is to transmit $X_c = X_c^{(1)} \bigoplus X_c^{(2)}$, this choice produces the outputs

$$Y_1 = \boldsymbol{S}^{m-n_{11}} X_1 + \boldsymbol{S}^{m-n_{1c}} X_c^{(2)}$$
$$Y_2 = \boldsymbol{S}^{m-n_{22}} X_2 + \boldsymbol{S}^{m-n_{2c}} X_c^{(1)}.$$

The achievability of $R_1 = n_{11}$ is proved by showing that it possible to recover $X_1$ from $Y_1$. We can rewrite $Y_1$ as

$$Y_1 = [\boldsymbol{0}^{m-n_{11}}\ \boldsymbol{b}_1^{n_{11}}] \oplus [\boldsymbol{0}^{m-n_{1c}+n_{2c}-n_{21}}\ (\boldsymbol{b}_1)_{n_{11}-[n_{21}-[n_{1c}-n_{2c}]^+]^+}^{n_{11}}\ \boldsymbol{0}^{[n_{2c}-n_{1c}]^+}]^T.$$

Decoding $X_1$ from $Y_1$ corresponds to solving the equation above for $\boldsymbol{b}_1$. The first $m - n_{11}$ equations are of the form $0 = 0$ and can be dropped. The remaining set of equations form a system of equations of $n_{11}$ equations in $n_{11}$ unknowns. The solution of this system of equations exists if and only if no bits in $\boldsymbol{b}_1$ both in $X_1$ and in $X_c^{(2)}$ align at the same level. In fact if one bit aligns at the same level in $X_1$ and in $X_c^{(2)}$ we have $(\boldsymbol{b}_1)_i \oplus (\boldsymbol{b}_1)_i = 0$ regardless of the value of $(\boldsymbol{b}_1)_i$ and thus the value of $(\boldsymbol{b}_1)_i$ cannot be determined from $Y_1$. Since $X_c^{(2)}$ is a down-shifted version $X_1$, this can be guaranteed by avoiding that $(\boldsymbol{b}_1)_{n_{11}}$ align at the same level, that is:

$$n_{11} \neq [n_{21} - [n_{1c} - n_{2c}]^+]^+,$$

but this is always verified since $n_{11} > n_{21}$.

By symmetric arguments, we may show the achievability of $R_2 = n_{22}$ under the condition $n_{22} > n_{21}$. This completes the achievability proof.

### E. Proof of Corollary IV.2

When $n_{12} = n_{21} = 0$ the outer bound of Th. III.3 can be reduced to

$$R_1 \leq \max\{n_{11}, n_{1c}\} \tag{8a}$$

$$R_2 \leq \max\{n_{22}, n_{2c}\} \tag{8b}$$

$$R_1 + R_2 \leq 1_{\{n_{11} \neq n_{1c} - n_{2c}\}} \left([n_{11} - n_{1c}]^+ + \max\{n_{22} + n_{1c}, n_{2c}\}\right)$$
$$+ 1_{\{n_{11} = n_{1c} - n_{2c}\}} (\max\{n_{22}, n_{2c}\} + n_{11}) \tag{8c}$$

$$R_1 + R_2 \leq 1_{\{n_{22} \neq n_{2c} - n_{1c}\}} \left([n_{22} - n_{2c}]^+ + \max\{n_{11} + n_{2c}, n_{1c}\}\right)$$
$$+ 1_{\{n_{22} = n_{2c} - n_{1c}\}} (\max\{n_{11}, n_{1c}\} + n_{22}) \tag{8d}$$

$$R_1 + R_2 \leq \max\{n_{11}, n_{1c}\} + \max\{n_{22}, n_{2c}\} \tag{8e}$$

$$2R_1 + R_2 \leq \max\{n_{11}, n_{1c}\}$$
$$+ \max\{n_{11}, , n_{1c}\} + \max\{n_{22}, n_{2c}\} \tag{8f}$$

$$R_1 + 2R_2 \leq \max\{n_{22}, n_{2c}\}$$
$$+ \max\{n_{11}, n_{1c}\} + \max\{n_{22}, n_{2c}\}. \tag{8g}$$

Since (8f) = (8a) + (8e) and (8g) = (8b) + (8e), we can drop these bounds. Also

$$\max\{n_{11}, n_{1c}\} + \max\{n_{22}, n_{2c}\} \geq \max\{n_{11}, n_{1c}\} + n_{22}$$
$$\max\{n_{11}, n_{1c}\} + \max\{n_{22}, n_{2c}\} \geq \max\{n_{11}, n_{1c}\} + n_{22}$$

and

$$\max\{n_{11}, n_{1c}\} + \max\{n_{22}, n_{2c}\} \geq \max\{n_{22} + n_{1c}, n_{2c}\}$$
$$\max\{n_{11}, n_{1c}\} + \max\{n_{22}, n_{2c}\} \geq \max\{n_{11} + n_{2c}, n_{1c}\}$$

so we can rewrite the outer bound as

$$R_1 \leq \max\{n_{11}, n_{1c}\} \tag{9a}$$

$$R_2 \leq \max\{n_{22}, n_{2c}\} \tag{9b}$$

$$R_1 + R_2 \leq 1_{\{n_{11} \neq n_{1c} - n_{2c}\}} \left([n_{11} - n_{1c}]^+ + \max\{n_{22} + n_{1c}, n_{2c}\}\right)$$
$$+ 1_{\{n_{11} = n_{1c} - n_{2c}\}} (\max\{n_{22}, n_{2c}\} + n_{11}) \tag{9c}$$

$$R_1 + R_2 \leq 1_{\{n_{22} \neq n_{2c} - n_{1c}\}} \left([n_{22} - n_{2c}]^+ + \max\{n_{11} + n_{2c}, n_{1c}\}\right)$$
$$+ 1_{\{n_{22} = n_{2c} - n_{1c}\}} (\max\{n_{11}, n_{1c}\} + n_{22}) \tag{9d}$$

Equations (9c) and (9d) contain an indicator function that depends on the condition $n_{11} = n_{1c} - n_{2c}$ and $n_{22} = n_{2c} - n_{1c}$ respectively. These conditions may be considered to be "degenerate" conditions as they imply a loss of degrees of freedom in the expression of $(Y_1, Y_2)$ as a function of the inputs $X_i$, $\in \{1, c, 2\}$. Under these conditions one input $X_k$, $k \in \{1, 2\}$ and $X_c$ align at both decoders in the same way so that they appear to be a single input $X_k \oplus X_c$. For this reason we must consider these conditions are separately.

The proof develops as follows:

i) achievability for $n_{22} = n_{2c} - n_{1c}$ and $n_{11} = n_{1c} - n_{2c}$,
ii) achievability for for $n_{22} \neq n_{2c} - n_{1c}$ and $n_{11} = n_{1c} - n_{2c}$, and by symmetry for $n_{22} = n_{2c} - n_{1c}$ and $n_{11} \neq n_{1c} - n_{2c}$
iii) achievability for $n_{22} \neq n_{2c} - n_{1c}$ and $n_{11} \neq n_{1c} - n_{2c}$

**Case i)** $n_{22} = n_{2c} - n_{1c}$ **and** $n_{11} = n_{1c} - n_{2c}$: then $n_{1c} = n_{2c} = n_c$ and $n_{11} = n_{22} = 0$. This means that the channel reduces to a degenerate broadcast channel with $Y_1 = Y_2$. The outer bound expression reduces to

$$R_1 \leq n_c \tag{10a}$$
$$R_2 \leq n_c \tag{10b}$$
$$R_1 + R_2 \leq n_c \tag{10c}$$
$$\tag{10d}$$

which in turn reduces to

$$R_1 + R_2 \leq n_c$$

which is always trivially achievable.

**Case ii)** $n_{11} = n_{1c} - n_{2c}$ **and** $n_{22} \neq n_{2c} - n_{1c}$: In this case the outer bound reduces to

$$R_1 \leq n_{1c} \tag{11a}$$
$$R_2 \leq \max\{n_{22}, n_{2c}\} \tag{11b}$$
$$R_1 + R_2 \leq n_{1c} + [n_{22} - n_{2c}]^+ \tag{11c}$$
$$R_1 + R_2 \leq n_{1c} + [n_{22} - n_{2c}]^+ \tag{11d}$$
$$\tag{11e}$$

Equation (11d) is redundant and the outer bound expression simplifies to

$$R_1 \leq n_{1c} \tag{12a}$$
$$R_2 \leq \max\{n_{22}, n_{2c}\} \tag{12b}$$
$$R_1 + R_2 \leq n_{1c} + [n_{22} - n_{2c}]^+. \tag{12c}$$

This region has corner points

$$A = (R_1^A, R_2^A) = (n_{1c}, [n_{22} - n_{2c}]^+)$$
$$B = (R_1^B, R_2^B) = (n_{1c} - n_{2c}, \max\{n_{22}, n_{2c}\}).$$

*Achieving corner point A:* consider the following scheme:
- Transmitter 1 (TX 1) transmits $n_{11}$ bits along its direct link to receiver 1 (RX 1) as $X_1 = [\boldsymbol{b}_1^{n_{11}} \ \mathbf{0}^{m-n_{11}}]^T$.
- Transmitter 2 (TX 2) transmits $n_{22}$ bits along its direct link to receiver 2 (RX 2) as $X_2 = [\boldsymbol{b}_2^{n_{22}} \ \mathbf{0}^{m-n_{22}}]^T$.
- The cognitive relay (CR) sends $n_{1c} - n_{11}$ to RX 1 to achieve the full rate $R_1 = n_{11}$. Since $n_{1c} > n_{2c}$, this set of bits interfere with the transmission of TX 2. The cognitive relay sends $X_c = [\boldsymbol{b}_1^{n_{1c}-n_{11}} \ \mathbf{0}^{m-(n_{1c}-n_{11})}]^T$.

The rate of user 1 is $R_1 = n_{1c}$ since at each channel transmission $n_{11}$ bits are received from TX 1 and $n_{1c} - n_{11}$ from CR. RX 2 receives $n_{22}$ bits frow TX 2 but $n_{1c} - n_{11} = n_{2c}$ bits of interference from the CR, therefore achieving rate $[n_{22} - n_{2c}]^+$

*Achieving corner point B:* we utilize a similar scheme, and let $X_1$ and $X_2$ be defined as in the achievability scheme for corner point $A$. However, the CR sends $[n_{2c} - n_{22}]^+$ to RX 2 to achieve the full rate $R_2 = n_{22}$. Since $n_{1c} > n_{2c}$, this set of bits interfere with the transmission of TX 1. The cognitive relay sends $X_c = [\boldsymbol{b}_2^{[n_{2c}-n_{22}]^+} \ \mathbf{0}^{m-[n_{2c}-n_{22}]^+}]^T$.

The rate of user 2 is $R_2 = \max\{n_{22}, n_{2c}\}$ since at each channel transmission $n_{22}$ bits are received from TX 2 and $[n_{2c} - n_{22}]^+$ from CR. RX 1 receives $n_{11}$ bits frow TX 1 and $[n_{2c} - n_{22}]^+$ bits of interference from the CR. This interference is always above the signal received from RX 1 since

$$n_{1c} - [n_{2c} - n_{22}]^+ \geq n_{1c} - n_{2c} = n_{11},$$

so $R_1 = n_{11}$.

The case $n_{11} \geq n_{1c} - n_{2c}$ and $n_{22} = n_{2c} - n_{1c}$ is obtained from this proof by the channel symmetry.

**Case iii)** $n_{22} \neq n_{2c} - n_{1c}$ **and** $n_{11} \neq n_{1c} - n_{2c}$: As pointed out in Section IV-B, we divide the proof into three regions:
- *weak cognition at both decoders:* $n_{11} \geq n_{1c}$, $n_{22} \geq n_{2c}$,
- *strong cognition at both decoders:* $n_{11} < n_{1c}$, $n_{22} < n_{2c}$,
- *strong cognition at one decoder and weak cognition at the other:* $n_{11} \geq n_{1c}$, $n_{22} < n_{2c}$.

We present here the achievability for the first two cases, the remaining case is presented in Section IV-B.

*1) Capacity for weak cognition at both decoders:* Under these conditions the outer bound of (9) reduces to

$$\begin{aligned} R_1 &\leq n_{11} \\ R_2 &\leq n_{22} \\ R_1 + R_2 &\leq n_{11} + n_{22} \\ R_1 + R_2 &\leq n_{22} + n_{11} \end{aligned}$$

which is equivalent to

$$\begin{aligned} R_1 &\leq n_{11} \\ R_2 &\leq n_{22}. \end{aligned}$$

In this case the outer bound can be achieved by ignoring the cognitive relay. Each encoder transmits along the direct link while the cognitive relay is silent, Let the channel input be

$$\begin{aligned} X_1 &= [\boldsymbol{b}^{n_{11}} \ \boldsymbol{0}^{m-n_{11}}]^T \\ X_2 &= [\boldsymbol{b}^{n_{22}} \ \boldsymbol{0}^{m-n_{22}}]^T, \end{aligned}$$

which trivially achieves the rate point $(R_1, R_2) = (n_{11}, n_{22})$. and is sufficient to show the achievability of the full outer bound region.

*2) Capacity for strong cognition at both decoders:* Under these conditions the outer bound of (9) becomes

$$\begin{aligned} R_1 &\leq n_{1c} \\ R_2 &\leq n_{2c} \\ R_1 + R_2 &\leq \max\{n_{22} + n_{1c}, n_{2c}\} \\ R_1 + R_2 &\leq \max\{n_{11} + n_{2c}, n_{1c}\} \end{aligned}$$

This region can be shown to be achievable by showing the achievability of the two corners where the sum rate intersects a bound on the single rates $R_1$ or $R_2$. The points $(n_{1c}, 0)$ and $(0, n_{2c})$ are trivially achievable by letting one of the transmitters be silent. Given the symmetry of the outer bound, is sufficient to prove the achievability of one corner of the outer bound. The achievability of the other corner is then obtained by switching the role of the users. We focus on the corner point where the maximal $R_1$ meets the sum rate outer bound, that is:

$$A = (R_1^A, R_2^A) = (n_{1c}, \min\{\max\{n_{22}, [n_{2c} - n_{1c}]\}, n_{2c} - (n_{1c} - n_{11})\}).$$

This point may be achieved by having cognitive user help TX 1 achieve $R_1 = n_{1c}$ whilst minimizing the interference at receiver 2 as follows:

- TX 1 transmits $n_{11}$ bits for RX 1 through the direct link as $X_1 = [\boldsymbol{b}_1^{n_{11}} \ \boldsymbol{0}^{m-n_{11}}]^T$
- the CR transmits $n_{1c} - n_{11}$ bits to RX 1 to achieve the full rate $R_1$
- the CR transmits $[n_{2c} - n_{1c}]^+$ bits for RX 2 without creating interference at RX 1 as

$$X_c = [(\boldsymbol{b}_1)_{n_{11}}^{n_{1c}} \ \boldsymbol{0}^{n_{11}} \ \boldsymbol{b}_2^{[n_{2c}-n_{1c}]^+} \ \boldsymbol{0}^{m-\max\{n_{1c}, n_{2c}\}}]^T$$

- TX 2 transmits $\min\{n_{11}, n_{22} - [n_{2c} - n_{1c}]^+\}$ to be received above the bits broadcasted from the CR at RX 2 as

$$X_2 = [(\boldsymbol{b}_2)_{[n_{2c}-n_{1c}]^+}^{[n_{2c}-n_{1c}]^+ + \min\{n_{11}, n_{22}-[n_{2c}-n_{1c}]^+\}} \ \boldsymbol{0}^{[n_{2c}-n_{1c}]^+} \ \boldsymbol{0}^{m-n_{22}}]^T.$$

RX 1 decodes $n_{1c}$ bits in total: $n_{11}$ from TX 1 and the remaining ones from the CR. RX 2 receives all the bits that the CR can allocate to it without creating interference at RX 1 plus the bits that be allocated on the direct link $n_{22}$. The interference at RX 2 is received from the level $n_{2c} - (n_{1c} - n_{11})$ to the level $n_{2c}$. The transmitted bits from CR to RX 2 can be at most $[n_{2c} - n_{1c}]$ and from TX 2 to RX 2 at most $n_{22}$. From this we conclude that the achieved $R_2$ indeed corresponds to the corner point $A$ since $R_2 = \min\{\max\{n_{22}, [n_{2c} - n_{1c}]^+\}, n_{2c} - (n_{1c} - n_{11})\}$.